\documentclass[fleqn,usenatbib]{mnras}

\usepackage{newtxtext,newtxmath}

\usepackage[T1]{fontenc}

\DeclareRobustCommand{\VAN}[3]{#2}
\let\VANthebibliography\thebibliography
\def\thebibliography{\DeclareRobustCommand{\VAN}[3]{##3}\VANthebibliography}

\newcommand{\Msun}{\rm{M}_{\odot}}

\usepackage{graphicx}	
\usepackage{amsmath}	

\title[On $N$-body simulations of globular cluster streams]{On $N$-body simulations of globular cluster streams}

\author[Banik \& Bovy]{
Nilanjan Banik$^{1}$\thanks{E-mail: banik@tamu.edu} \&
Jo Bovy$^{2}$
\\
$^{1}$Mitchell Institute for Fundamental Physics and Astronomy, Department of Physics and Astronomy, Texas A\&M University, College Station, TX 77843, USA\\
$^{2}$ Department of Astronomy and Astrophysics, University of Toronto, 50 St. George Street, Toronto, ON, M5S 3H4, Canada \\
}

\date{Accepted XXX. Received YYY; in original form ZZZ}

\begin{document}
\label{firstpage}
\pagerange{\pageref{firstpage}--\pageref{lastpage}}
\maketitle

\begin{abstract}
Stellar tidal streams are sensitive tracers of the properties of the gravitational potential in which they orbit and detailed observations of their density structure can be used to place stringent constraints on fluctuations in the potential caused by, e.g., the expected populations of dark matter subhalos in the standard cold dark matter paradigm (CDM). Simulations of the evolution of stellar streams in live $N$-body halos without low-mass dark-matter subhalos, however, indicate that streams exhibit significant perturbations on small scales even in the absence of substructure. Here we demonstrate, using high-resolution $N$-body simulations combined with sophisticated semi-analytic and simple analytic models, that the mass resolutions of $10^4$--$10^5\,\Msun$ commonly used to perform such simulations cause spurious stream density variations with a similar magnitude on large scales as those expected from a CDM-like subhalo population and an order of magnitude larger on small, yet observable, scales. We estimate that mass resolutions of $\approx100\,\Msun$ ($\approx1\,\Msun$) are necessary for spurious, numerical density variations to be well below the CDM subhalo expectation on large (small) scales. That streams are sensitive to a simulation's particle mass down to such small masses indicates that streams are sensitive to dark matter clustering down to these low masses if a significant fraction of the dark matter is clustered or concentrated in this way, for example, in MACHO models with masses of $10$--$100\,\Msun$.
\end{abstract}




\section{Introduction}

The standard cold dark matter (CDM) picture of structure formation predicts that a Milky-Way-size galaxy halo should host numerous dark substructures (``subhalos'') with a range of masses down to many of orders of magnitudes below the typical dwarf galaxy mass \citep{Diemand2008,Springel2008}. Detecting these substructures can not only provide tell-tale evidence for the existence of dark matter but will also enable us to put stringent constraints on its particle nature. Unfortunately, being low in mass, these substructures are not able to initiate star formation and hence are undetectable directly by telescopes observing in the electromagnetic spectrum. 

Flux perturbations in strong gravitational lensing systems due to these substructures provides an indirect way of detecting and analyzing these low mass substructures \citep{Dalal2002,Vegetti:2008eg,Despali:2016meh,Gilman2019}. A powerful complementary approach is to study stellar density perturbations along stellar streams left as a result of gravitational encounters with these dark subhalos. Stellar streams emerging from tidally disrupting globular clusters in our Galaxy have a largely one-dimensional structure and in the absence of external perturbations have a fairly uniform stellar density along its length \citep{Johnston1998,Sanders2013,Bovy2014}. Owing to the very low velocity dispersion among its member stars, these streams are dynamically cold and hence are extremely sensitive to perturbations in the underlying gravitational potential. Close flybys of subhalos can therefore leave visible imprints in the stellar density along the streams in the form of gaps \citep{Ibata2001,Johnston2002,Siegal-Gaskins2008,Carlberg2009}. While these gaps can can be individually analyzed to infer the properties of the perturbing subhalo \citep{Yoon2011,Carlberg2012,Carlberg2013,Erkal2015,Erkal2015a,Sanders2016}, powerful statistical techniques based on the power spectrum of the full stream density \citep{Bovy2016a} have been applied to observed GD-1 and Pal 5 stream data to constrain the abundance of subhalos in the mass range $10^{6} - 10^{9} \ \Msun$ and constrain the particle mass of thermal dark matter \citep{Banik2020long,Banik2020letter}. 

A key step in effectively using streams as probes for dark matter substructure is to identify sources of stream density perturbations other than subhalo encounters and modeling their effects in the analysis. Depending on the orbit, stellar streams can be significantly perturbed by the baryonic structures in our Galaxy such as the bar \citep{Erkal2017,Pearson2017,Banik2019}, the spiral arms \citep{Banik2019} and the Giant Molecular Clouds (GMCs) \citep{Amorisco2016,Banik2019}.
Another source of stream density variations are the epicyclic overdensities which arise since stars are tidally ejected in bursts near the pericentric passage of the progenitor globular cluster \citep{Kuepper2009,Kuepper2011}. Such overdensities along the stream due to episodic tidal stripping were studied in \citep{Sanders2016,Bovy2016a} and were found to quickly disperse out as the ejected stars mixed due to their velocity dispersion. Stream regions near the progenitor that are dynamically young may still have such overdensities since the stars did not have sufficient time to mix, therefore such regions need to be judiciously excluded in inferring dark matter properties. 

Recently, \citet{Carlberg2017} and \citet{Carlberg2018}  investigated the dynamical evolution of globular clusters and their tidally disrupted streams in cosmological $N$-body simulations by placing the progenitor clusters on near circular orbits inside reconstituted subhalos from the Via-Lactea II simulation \citep{Madau2008} and evolving them. These subhalos along with their globular clusters and their tidally disrupted streams were eventually accreted onto the host halo and the final stream structures had a wide range of density variations. In order to test whether these density variations were caused by gravitational encounters with lower mass subhalos ($ \lesssim 4\times 10^{8} \ \Msun$), a similar simulation without these lower mass subhalos was run and the resulting stream densities still had similar variations. Based on this finding, it was concluded that the stream density variations were not caused by the low mass subhalo impacts. 

However, while \citet{Carlberg2017} was careful to use a mass resolution of $2\times 10^5\,\Msun$ such that \emph{heating} from the massive $N$-body particles is insignificant, it is unclear whether \emph{coherent} perturbations to the kinematics of simulated tidal streams by the $N$-body particles are significant. Coherent perturbations are what is relevant for determining whether stream density variations from numerical effects can be mistaken for true subhalo-induced variations, because the latter produce only coherent fluctuations without significant heating. Because $N$-body simulations are useful for understanding the evolution of tidal streams in the full cosmological context, in this paper, we investigate whether significant stream density variations are created in simulated globular-cluster streams evolved within a Milky Way like live halo of different mass resolutions ($10^{5}\,\Msun$, similar to \citealt{Carlberg2018}, and $10^{4}\,\Msun$). We show that in the absence of subhalos, numerically-induced stream density fluctuations at these mass resolutions exceed the expected subhalo signal by an order of magnitude and we derive the necessary mass resolution for numerical effects to be well below the expected CDM subhalo signal.

\section{Simulations}
\label{sec:method}

\begin{figure}
\includegraphics[width = 0.45\textwidth]{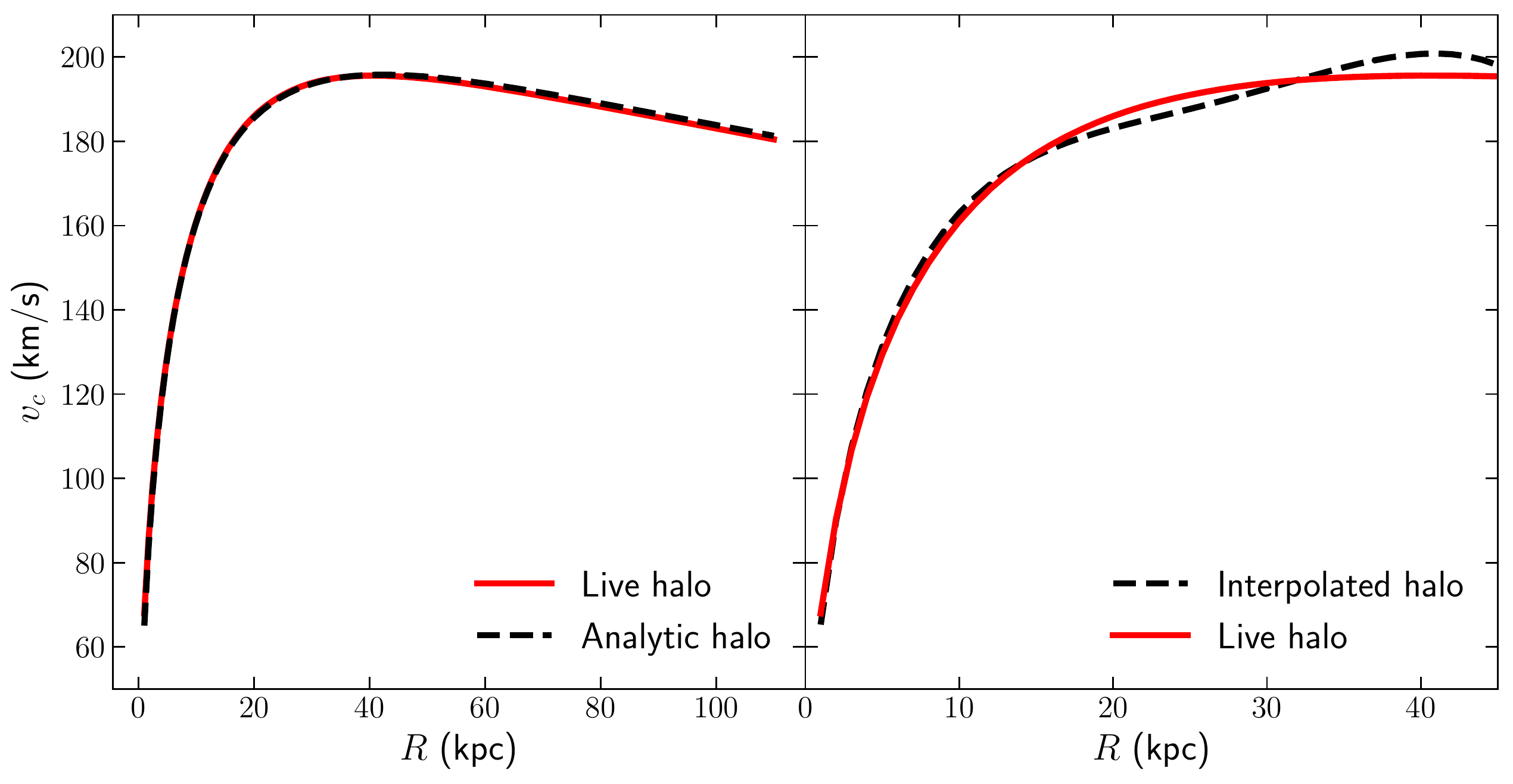}
\caption{Rotation curves of our live, analytic, and smooth halos. \textit{Left panel}: the solid red curve shows the rotation curve of the live $N$-body halo, which is initialized as an NFW profile and has a mass resolution of $10^{5} \ \Msun$, and the dashed black curve shows the analytic rotation curve with the same NFW profile parameters. \textit{Right panel}: the rotation curve of the same live halo is again shown as a solid red curve and the dashed black curve displays the rotation curve of the static, smooth interpolated representation of the same live halo as described in the text. All of the rotation curves agree well, indicating that the overall mass distribution of all of the models that we simulate is similar.}
\label{fig:vcirc}
\end{figure}

We run three different $N$-body simulations of the dynamical evolution of a globular cluster with an orbit that is similar to that of the GD-1 stream \citep{Grillmair2006} in the Milky Way, because GD-1 is currently the best stream to study subhalo-induced density perturbations \citep{Banik2020long}. We let the cluster and stream evolve for 4 Gyr, the approximate age of the GD-1 stream. The progenitor cluster is modeled by an isothermal King profile \citep{King1966} with total mass $7360\,\Msun$, a half-mass radius of 20 pc, and a dimensionless central potential depth $W_{0}$ is set to 7. We use \textsc{LIMEPY} \citep{Gieles2015} to generate the initial condition of the cluster with 100,000 equal-mass particles; because the GD-1 progenitor and its stars are likely $\approx12\,\mathrm{Gyr}$ old, the effects of stellar evolution and relaxation are small, allowing us to model the system as a collisionless system \citep{Webb2018}. The velocity dispersion of the initial cluster is $0.46\,\mathrm{km\,s}^{-1}$. The host potential consists of a Milky-Way-like halo and disk. The halo is modeled as an NFW profile \citep{Navarro97a} using the best fit parameters from \citep{McMillan2017} (Table 3), i.e. scale length $r_{h} = 19.6$ kpc and density parameter $\rho_{0} = 8.54\times 10^{6}\,\Msun\,\rm{kpc}^{-3}$. We generate two live Milky Way like halo initial conditions using \textsc{GalactICS} \citep{Kuijken1995,Widrow2005,Widrow2008,Deg2019} with particle mass resolution of $10^{5}\,\Msun$ and $10^{4}\,\Msun$. The left panel of Figure \ref{fig:vcirc} shows that the rotation curve of the live halo is consistent with that of the analytic NFW halo with the same parameters. The disk is modeled as a Miyamoto-Nagai profile with a total mass of $6.8\times 10^{10}\,\Msun$, a radial scale length of 3 kpc, and a scale height of 0.28 kpc \citep{Bovy2015}. Because we are mainly interested in the studying the effects of mass resolution of the halo on the globular cluster stream, we evaluate the disk potential analytically as an external force in the $N$-body simulations. To study the evolution of the globular cluster stream in a smooth halo similar to the $N$-body one, we determine the radial acceleration $a_{r}(r)$ by computing $-GM(<r)/r^{2}$ on a uniform radial grid using the positions of the live halo particles in the $10^{5} \ \Msun$ resolution case and interpolating them to create a spherical, static, and smooth representation of the live halo. The right panel of Figure \ref{fig:vcirc} compares the rotation curve of the live halo with that from the interpolated smooth halo within 50 kpc of the host center and shows good agreement. We limit our interpolation to 50 kpc, because the stream orbit is confined well within this radial range. To summarize: we run three main simulations: a live halo with mass resolution $10^5\,\Msun$, a live halo with $10^4\,\Msun$ resolution, and a static halo created from smoothing out the $10^{5} \ \Msun$ resolution live halo through interpolation.

\begin{figure}
\includegraphics[width =0.45\textwidth]{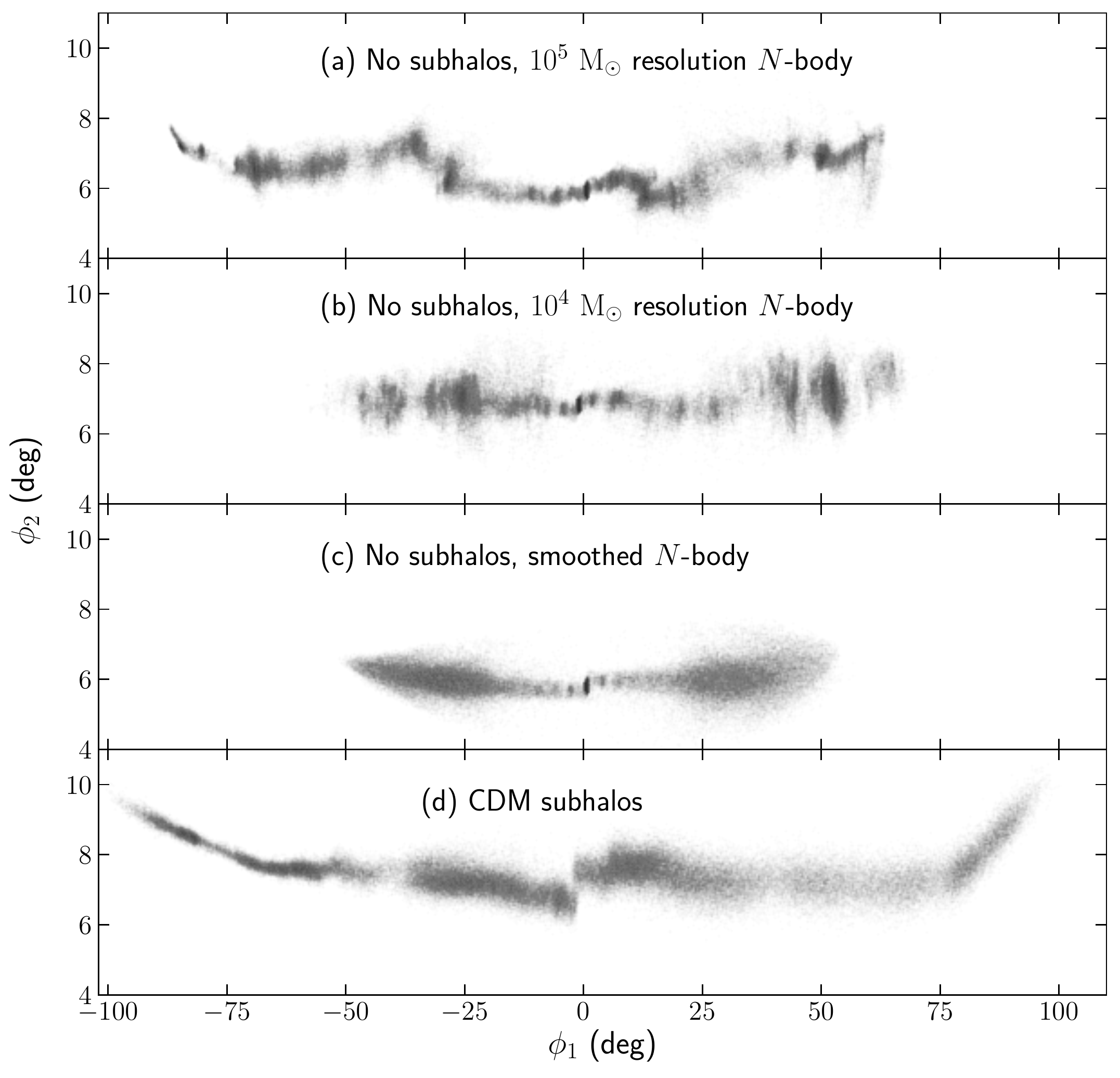}
\caption{The simulated streams after 4 Gyr of evolution represented in the $(\phi_{1},\phi_{2})$ angular sky coordinate frame centered at the progenitor location. \textit{Panel (a):} $N$-body run with live halo of mass resolution of $10^{5}\,\Msun$, \textit{panel (b):} $N$-body run with live halo of mass resolution of $10^{4}\,\Msun$, \textit{panel (c):} static, smoothed $N$-body halo, \textit{panel (d):} simulated stream in the \texttt{streampepperdf} framework that was impacted by the CDM abundance of subhalos in the mass range $10^{5}--10^{9}\,\Msun$. For the $N$-body runs, the progenitor can still be seen at $\phi_{1} \sim 0$. In both panels (a) and (b), significant density variations can be seen along the stream. In panel (c), epicylic overdensities can be clearly seen within $20^{\circ}$ around the progenitor. Overall, the streams evolved in the live $N$-body realization of a smooth halo show significantly more density variations than in the smooth or CDM-like halos.}
\label{fig:stream_sky_coord}
\end{figure}

\begin{figure}
\includegraphics[width = 0.45\textwidth]{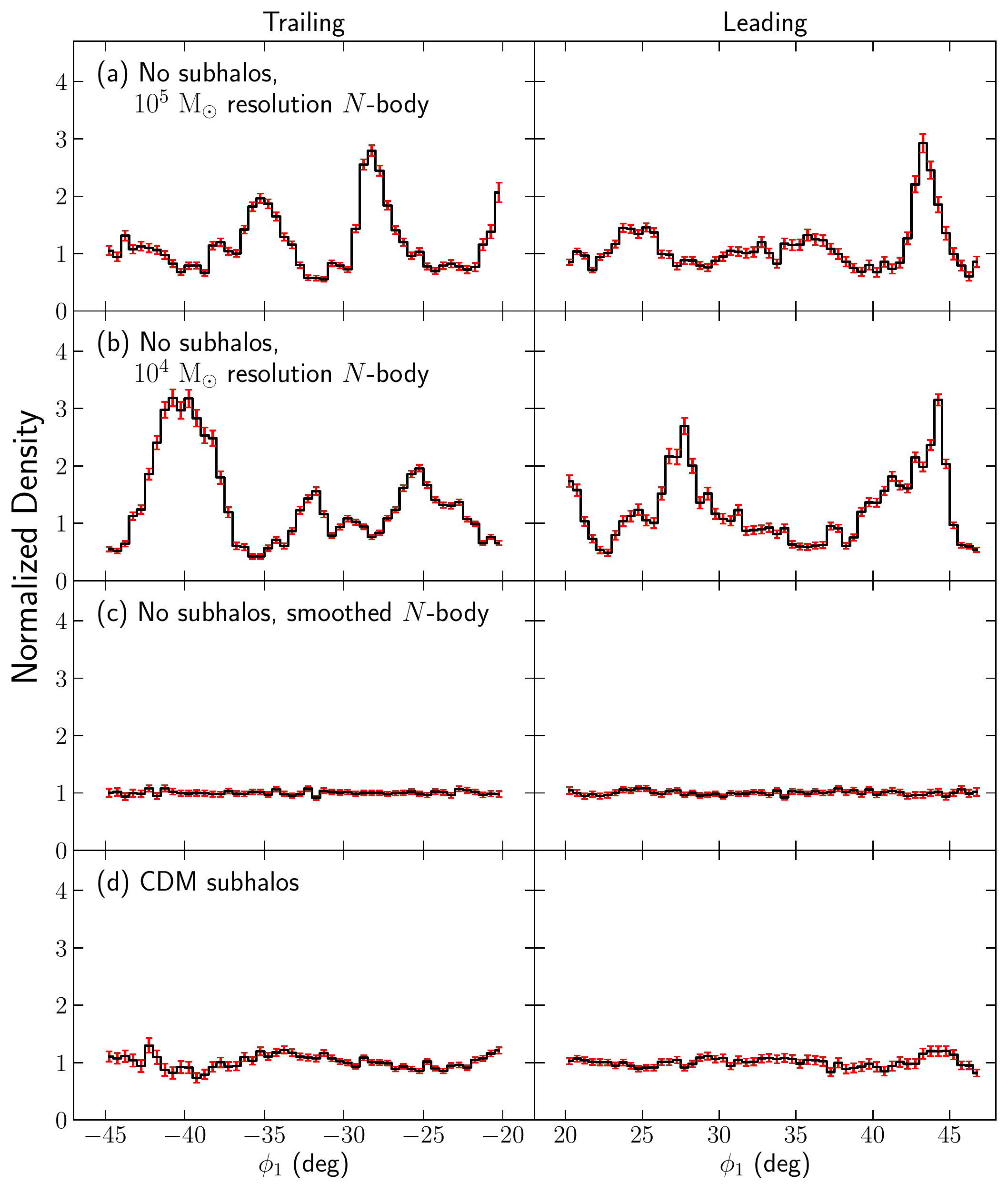}
\caption{Normalized stream density of the four cases shown in Figure \ref{fig:stream_sky_coord}. This is computed by first binning the particles along the stream in $0.5^{\circ}$ bins in $\phi_{1}$ and then dividing by a $3^{\rm{rd}}$ order smoothing polynomial fit. The right columns show the leading arm and the left columns show the trailing arm after excluding $20^{\circ}$ around the progenitor and considering $30^{\circ}$ and $25^{\circ}$ along the leading and trailing arm respectively. The red error bars are the shot noise in each bin.}
\label{fig:stream_dens}
\end{figure}

We obtain a GD-1 like orbit for the cluster for each of the three simulations that we run by placing our cluster at the assumed current phase space coordinates of the GD-1 progenitor following \citet{Webb2018}, flipping the sign of the velocity, and backwards-evolving for 4 Gyr using the $N$-body code \textsc{GIZMO} \citep{Hopkins2014} which is based on the \textsc{GADGET} code \citep{Springel_Gadget}. In all simulations, we use a force softening of 2 pc for the cluster particles and in the live-halo simulations we use a softening of 200 pc and 20 pc for the halo particles in the $10^{5}\,\Msun$ and $10^{4}\,\Msun$ resolution cases, respectively. At the end of the backwards-evolution, we determine the phase space coordinate of the cluster's center of density using \texttt{clustertools}\footnote{Available at \url{https://github.com/webbjj/clustertools}~.} and we then place the same initial cluster at this point, flip the sign of the velocity, and forward-evolve it for 4 Gyr. At the end of the backward-forward simulation, the resulting cluster center was found to match the present day GD-1 progenitor's phase space location that we started with very well and a realistic tidal stream of escaped cluster particles forms in each simulation. The mean Galactocentric radius of the orbit is $\sim 17$ kpc, with a mean orbital eccentricity of 0.2, and peri- and apogalacticon distances of $\sim 13$ kpc and $\sim 20$ kpc, respectively, similar to the observed GD-1 stream \citep{Webb2018}. Because the initial condition of the halo was generated in the absence of a disk, placing the analytic disk at the center of the halo breaks its state of equilibrium. Therefore, we implement conditions on the analytic disk such that the halo particles (particle type 1) do not feel the gravitational force from the disk, while the cluster particles (particle type 4) feel the combined force from the halo and the disk. We have also ignored adiabatic contraction of the halo which should not have any effect at the small scales of stream density variations.  

\section{Stream density variations in $N$-body, smooth, and CDM halos}
\label{sec:results}

\begin{figure}
\includegraphics[width = 0.45\textwidth]{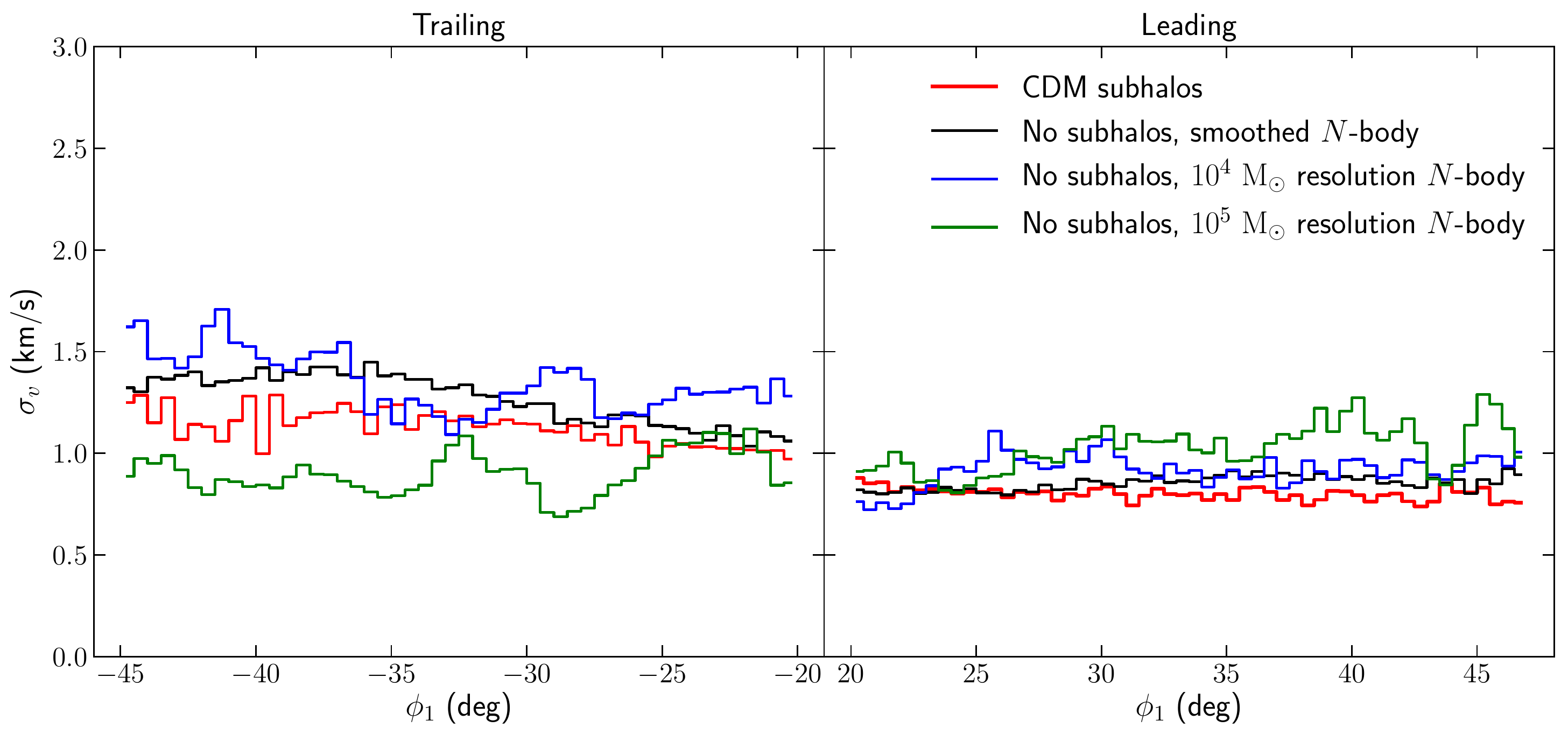}
\caption{Velocity dispersion along the stream in the four different cases shown in Figure \ref{fig:stream_sky_coord}. There is no substantial overall heating along the stream in any of the simulated cases.}
\label{fig:vel_disp}
\end{figure}

The resulting streams in all of our simulations are shown in panels (a),(b) and (c) in Figure \ref{fig:stream_sky_coord} where $(\phi_{1},\phi_{2})$ are coordinates along and transverse to the stream centered at the progenitor. The $(\phi_{1},\phi_{2})$ coordinate system is a rotation of the sky coordinate system that was determined by hand to align the stream to have constant $\phi_2 $ such that $\phi_1$ can be used as the along-stream angular coordinate.  In both live halo cases the resulting streams can be seen to have acquired substantial density variations that are clearly absent in the smoothed halo case. For reference, we show in panel (d) one realization of a similar stream that was impacted by a CDM abundance of subhalos in the mass range $10^{5} - 10^{9}\,\Msun$. This last case is generated using the stream-subhalo interaction framework \texttt{streampepperdf}\footnote{Available at \url{https://github.com/jobovy/streamgap-pepper}~.}, which is based on the \texttt{galpy} code \citep{Bovy2015}. The CDM subhalo abundance and their corresponding sizes were obtained following the fitting functions from \citet{Erkal2015a} that were based on the Aquarius simulations \citep{Springel2008}. The subhalos are set to impact the stream following the same steps as in \citet{Bovy2016a} and we refer the reader to that paper for full details.

\begin{figure*}
\includegraphics[width = 0.9\textwidth]{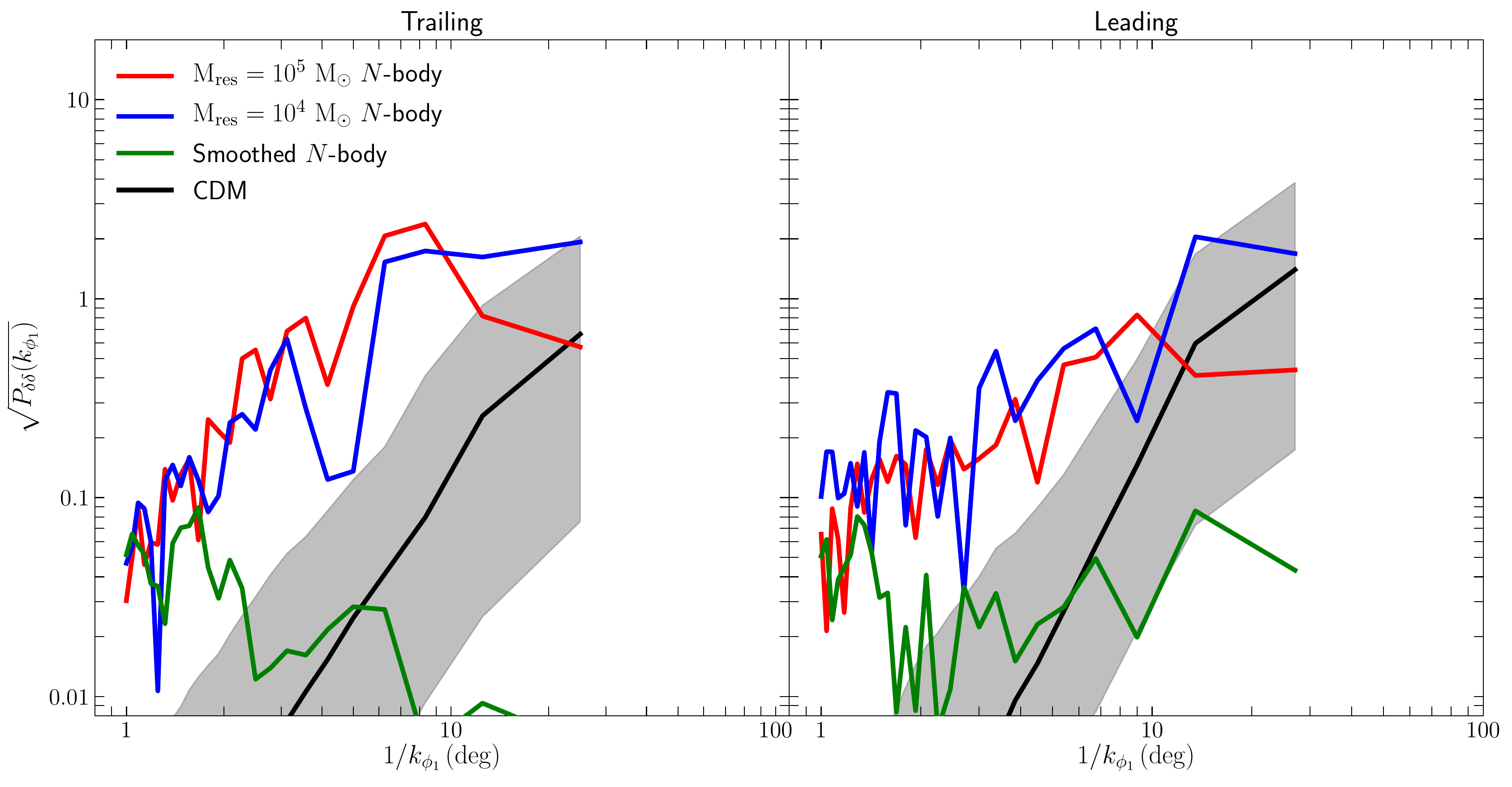}
\caption{Density power spectrum of the leading and trailing arm of streams simulated in live $N$-body halos with $10^5\,\Msun$ (red curve) and $10^4\,\Msun$ resolution. The green curve shows the power spectrum of the stream evolved in the static, smoothed $N$-body run. The black curve and the gray shaded region shows the median and $2\sigma$ dispersion of density power due to CDM subhalo impacts. At the largest scales the density power in the live halo cases are comparable to that due to impacts by CDM subhalos, while for the smoothed halo case the density power is substantially lower and consistent with noise. On small scales, the density power in the live halo cases exceed that expected from a CDM-like population of subhalos by an order of magnitude or more.}
\label{fig:Pk_live}
\end{figure*}
To quantify the amount of density variations in each case, we will compute the power spectrum of the normalized stream density following \citet{Bovy2016a}.  Variations in the stream density due to epicyclic pile ups can be most clearly seen within $\sim 20^{\circ}$ around the progenitor in the smoothed halo case, therefore we cut out that region from each case to exclude their contribution to the density variations. Furthermore, we restrict our analysis to $27^{\circ}$ and $25^{\circ}$  along the leading and trailing arms respectively for each case. This way we are comparing the density variations over the same angular range along the stream for the different cases. The densities are normalized by dividing out by a $3^{\rm{rd}}$ order smoothing polynomial fit following \citep{Bovy2016a}. The resulting normalized stream densities are shown in Figure \ref{fig:stream_dens}. As expected from Figure \ref{fig:stream_sky_coord}, the stream density in Figure \ref{fig:stream_dens} in the smoothed halo case is mostly flat, while the live halo cases display wide-scale density variations. The velocity dispersion along the stream length for the different cases are shown in Figure \ref{fig:vel_disp}, which demonstrates that heating of the less massive star particles of the cluster by the more massive dark matter particles of the halo is not significant, consistent with the analysis presented in \citep{Carlberg2017}.

The power spectrum of the stream density, computed using the same technique from \citep{Bovy2016a}, from the live and smoothed halo $N$-body runs are shown in Figure \ref{fig:Pk_live} by the red solid curve ($10^{5}\,\Msun$ resolution), the blue solid curve ($10^{4}\,\Msun$ resolution), and the green solid curve (smoothed static halo). Overplotted as the black solid line and gray shaded region are the median and $2\sigma$ dispersion in the stream density power respectively, of 1000 realizations of the stream impacted by a CDM abundance of subhalos. From this figure it is clear that at the largest angular scales, the density power accrued by the stream through the interactions with the halo particles in the $N$-body runs is comparable to that due to the CDM subhalo impacts. The power in the smoothed $N$-body halo case has comparatively very low power that is consistent with Poisson noise, which is $\approx 0.04$ for both arms. The power in both the live halo cases is at all scales much higher than the noise. This shows that even in the absence of subhalo impacts, interactions with the dark matter halo particles of resolution $10^{5}\,\Msun$ or $10^{4}\,\Msun$ can substantially perturb the stream density. 

\begin{figure*}
\includegraphics[width = 0.9\textwidth]{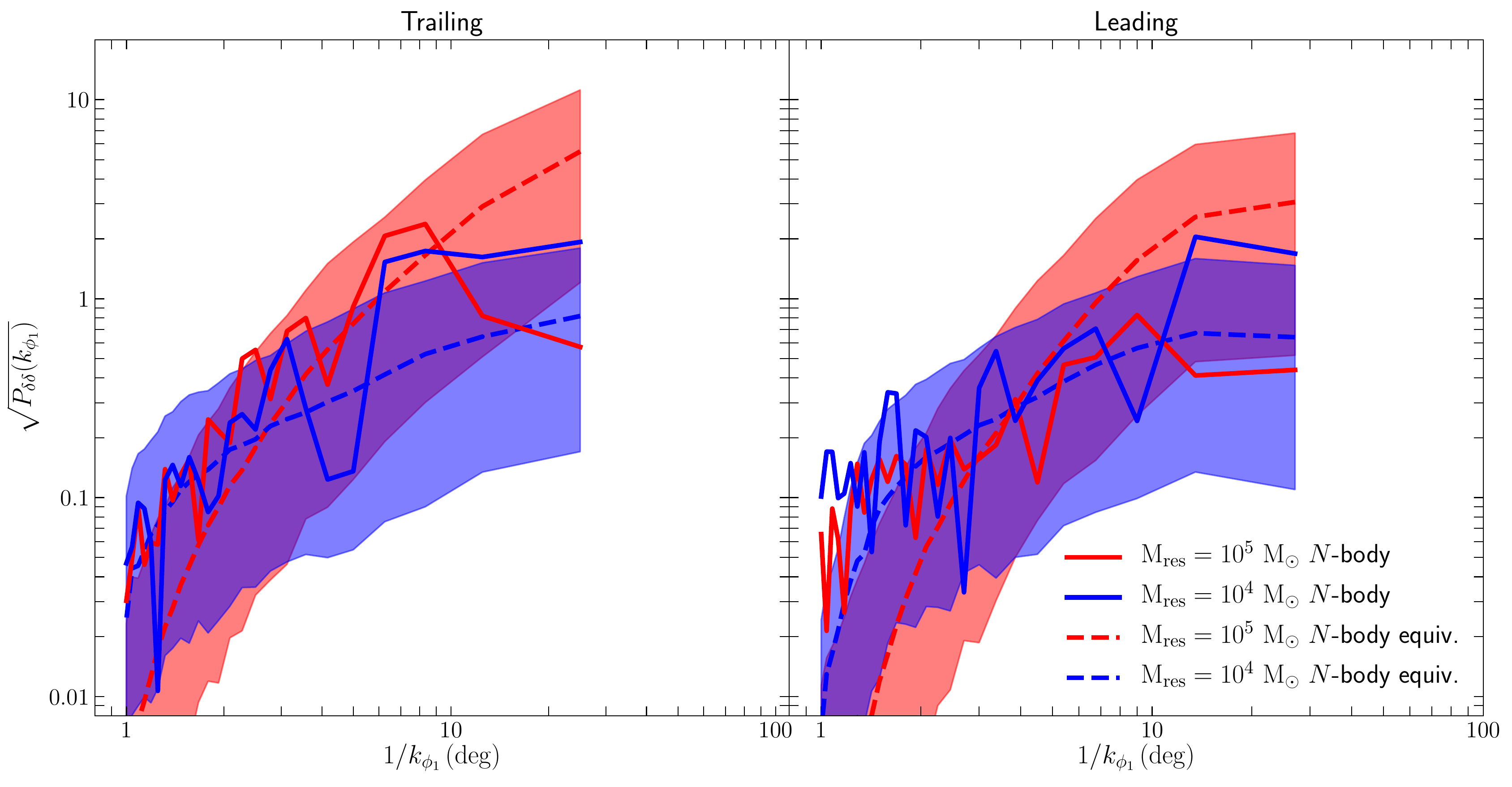}
\caption{Power spectrum of the leading and trailing arm in the live $N$-body runs (solid curves) compared with the case where the stream was impacted by a perturber population that has a similar abundance as the $N$-body particles and for which the perturbers have the same mass and similar structure as the softened $N$-body particles. For the latter case, the median (dashed curve) and $2\sigma$ dispersion (shaded region) are shown. The red color is used for the $10^5\,\Msun$ case while the blue color for the $10^4\,\Msun$ case. The good agreement between the $N$-body results and the perturber-population prediction demonstrates that the mass resolution is the dominant reason for the large density variations in the $N$-body streams in Figure \ref{fig:stream_sky_coord}.} 
\label{fig:Pk_Nbody}
\end{figure*}

Next, to conclusively demonstrate that these stream density variations are really due to interactions with the $N$-body particles, we evolve the same stream in the smoothed halo potential while impacting it with a population of subhalos similar in abundance and structure as the $N$-body particles. We want to check whether the resulting stream density power spectrum is comparable to those obtained in the live halo cases. To do this, we again use the fast stream-subhalo interaction simulator \texttt{streampepperdf} and model the subhalos as Plummer spheres with mass equal to the $N$-body particle mass and scale length equal to the Plummer equivalent softening of the $N$-body particles. The latter in \textsc{GIZMO} is $1/2.8$ times the force softening (see above). The expected number of subhalo encounters is directly proportional to its number density within the mean Galactocentric radius of the stream which is $\sim 17$ kpc in our case. The number density of $N$-body particles in the $10^{5} \ \Msun$ resolution case is around 2513 times the CDM predicted subhalo number density in the mass range $10^{4.5} - 10^{5.5}\,\Msun$ and for the $10^{4}\,\Msun$ resolution case this factor $n_{N\rm{-body}}/n_{\rm{subhalo}}$ is around 3168 times the CDM predicted subhalo number density in the mass range $10^{3.5} - 10^{4.5}\,\Msun$. Simulating all encounters from such a huge number of subhalos is computationally very challenging. Therefore we use an approximate method in which we evolve the stream with $10\times$ the CDM predicted subhalo number density and scale the resulting density power spectrum by $n_{N\rm{-body}}/n_{\rm{subhalo}}/10$, because uncorrelated subhalo encounters should contribute additively to the power spectrum (note that what we plot in Figure \ref{fig:Pk_live} is the square root of the power spectrum). We ran 1000 simulations each for the $10^{5}\,\Msun$ and $10^{4}\,\Msun$ resolution cases and the results are shown in Figure \ref{fig:Pk_Nbody}. The red and blue dashed curves show the median power in the $10^{5}\,\Msun$ and $10^{4}\,\Msun$ resolution cases, respectively. The $2\sigma$ dispersion of power in the corresponding cases is shown by the shaded regions. The density power from the live halo cases (solid curves) are consistent with the corresponding predicted density powers within their dispersion confirming that the density variations in the live $N$-body simulations are artefacts of the dark matter particle resolution. It can also be seen that for both these cases as well as for the live halo cases, there is much more power at small scales compared to the CDM case, which is expected since as shown in \citet{Bovy2016a}, low mass (such as $ \sim 10^{5}\,\Msun$) perturbers result in small scale density fluctuations. That said, we also find enormous power at large scales along the stream in these cases which seems counter-intuitive. This is because density fluctuations inflicted on streams grow linearly with time \citep{Erkal2015} and so over the course of its evolution the stream density structure is a combination of older large scale fluctuations and recent small scale fluctuations. 

\section{Discussion and Conclusions}
\label{sec:discussion}

We have investigated the effects of dark-matter particle mass resolution in $N$-body simulations on the evolution of globular cluster streams. Comparing $N$-body simulations of the evolution of a globular cluster and its tidal stream in live Milky Way like halos with resolutions of $10^{4}\,\Msun$ and $10^{5}\,\Msun$ with the same stream evolved in a static, smoothed dark matter halo demonstrates that, despite the absence of significant overall heating along the stream, in the live halo simulations the final stream density displays significant variations on all angular scales along the stream. In particular, the density power at the largest scales is comparable to that expected due to impacts by a CDM-like population of subhalos in the mass range $10^{5}$--$10^{9}\,\Msun$. We further verify that the stream density variations are due to interactions with the massive $N$-body particles by evolving the stream in the smoothed halo while impacting it with a population of subhalos similar in abundance and structure as the $N$-body particles. We found that the resulting density power is in agreement with that of the live halo cases. This conclusively shows that $N$-body resolutions of $10^{4}\,\Msun$ and $10^{5}\,\Msun$ can severely perturb globular cluster stream densities and that these mass resolutions---while high by present-day standards---are inadequate for studying the expected density variations along tidal streams in $N$-body simulations. In particular, the results presented here demonstrate that the density variations found in the cosmological simulations of \citet{Carlberg2018} in the absence of subhalo impacts are likely a resolution artifact owing to using a dark matter resolution of $2\times10^{5}\,\Msun$.

While we have used $N$-body simulations to determine the expected power spectrum of stream density variations in the presence of a population of massive $N$-body particles, when the entire halo is made up off $N = M_\mathrm{halo}/M_{\mathrm{res}}$ particles with mass $M_{\mathrm{res}}$, the number of impacts on a stream is high enough that we can estimate the induced power using simple statistical calculations. Following Appendix B of \citet{Dalal20a}, who work out the expected stream density power spectrum for a population of CDM subhalos, we can determine the approximate scaling of the power spectrum with mass. As shown by \citet{Dalal20a}, for perturbers of a single mass $M_{\mathrm{res}}$ with a spatial density $\bar{n}$, the expected power spectrum is $P_{\delta\delta} \propto M_{\mathrm{res}}^2\,\bar{n}$. When the entire halo consists of substructure of mass $m$, then $\bar{n} = \rho / M_{\mathrm{res}}$, where $\rho$ is the dark matter density and $P_{\delta\delta} \propto M_{\mathrm{res}}$ (where we drop the $\rho$ dependence because it does not change with resolution). This scaling holds down to angular scales $\theta \approx R/r$, where $R$ is the size of the perturber and $r$ is the Galactocentric radius (that is, $\theta$ is the angular size of perturbers as seen from the Galactic center); below this scale the power spectrum drops to zero quickly. Figures \ref{fig:Pk_live} and \ref{fig:Pk_Nbody} show that on large scales indeed approximately $P_{\delta\delta} \propto M_{\mathrm{res}}$. To be able to simulate the expected large-scale power without numerical artefacts requires at least an order of magnitude improvement in the $N$-body $\sqrt{P_{\delta\delta}}$ or a decrease in $N$-body particle mass of 100 (to $M_{\mathrm{res}} \approx 100\,\Msun$) and ideally smaller to avoid statistical upwards fluctuations in the numerical power. Because the power spectrum of artificial fluctuations declines much less steeply than that predicted by CDM subhalos, the resolution requirement becomes even more stringent at small scales; on the smallest scales at which the power is observable with future data (few degree; \citealt{Bovy2016a}) the numerical power $\sqrt{P_{\delta\delta}}$ is an order of magnitude larger than the predicted CDM power and a mass resolution of $1\,\Msun$ would be necessary for artefacts to be negligible. Such mass resolutions are higher than typically used, but are achievable with some current algorithms making efficient use of modern high-performance computing such as GPUs (e.g., \citealt{2014hpcn.conf...54B,2020MNRAS.499.2416A}).

Our results confirm that stellar streams in the Milky Way are extremely sensitive to fluctuations in the gravitational potential. While, as we have shown, this puts stringent constraints on $N$-body simulations of the evolution of such streams, it also presents an opportunity in that it further confirms that streams present one of the most sensitive ways to learn about non-standard dark matter models. In particular, our results indicate that streams are sensitive to dark matter clustering on mass scales of $\approx 10$--$100\,\Msun$ if a significant fraction of the dark matter participates in this clustering. One such dark-matter candidate is MACHOs (e.g., primordial black holes), where next-generation stream-density measurements on small scales by LSST at the Vera Rubin Observatory \citep{LSST} or the Roman Space Telescope \citep{RomanST} may be able to constrain MACHO dark matter in the interesting $\approx 10$--$100\,\Msun$ range \citep{Bird16a,Brandt16a,PBH}.

\section*{Acknowledgements}
We thank Jeremy Webb and Nathan Deg for useful discussions. NB thanks Nathan Deg for help in setting up and running \textsc{GalactICS}. JB received financial support from NSERC (funding reference number RGPIN-2020-04712), an Ontario Early Researcher Award (ER16-12-061), and from the Canada Research Chair program.
Portions of this research were conducted with high performance research computing resources provided by Texas A\&M University (\url{https://hprc.tamu.edu}).
\section*{Data Availability}
No new data were generated in support of this research.



\bibliographystyle{mnras}
\bibliography{Effect_of_Nbody_GlobClus_Stream} 








\bsp	
\label{lastpage}
\end{document}